\documentclass{webofc}
\usepackage[varg]{txfonts}   
\usepackage{svg}
\def\av#1{\langle #1 \rangle}

\begin{document}
\title{Strongly intensive fluctuations and correlations in  $\text{ultrarelativistic}$ nuclear collisions in the model with string fusion}
%
%

\author{\firstname{Vladimir} \lastname{Kovalenko}\inst{1}\fnsep\thanks{\email{v.kovalenko@spbu.ru}}
}

\institute{Saint Petersburg State University, Saint Petersburg, Russia
          }

\abstract{%
The several types of strongly intensive correlation variables are studied in nuclear collisions at LHC energy. These quantities are expected not to depend on centrality class width. They have been calculated in the dipole-based parton-string Monte Carlo model with string fusion. The centrality dependence of the mean transverse momentum correlation coefficient and strongly intensive quantity $\Sigma$ between multiplicity and $P_T$ have been obtained. Dynamical charge fluctuation $\nu_{dyn}$ has been also calculated and compared with experimental data. It is shown that string fusion improves agreement with the experiment.
}
\maketitle\vspace{-0.55cm}
\section{Introduction}
\label{intro}
The exploration of the QCD phase diagram and a search for new phenomena is one of the main subjects of heavy ion physics. The experimental investigation of the phase diagram of a strongly interacting matter is related to the study of the fluctuations and correlations of observables in the nuclear collisions at high energy\cite{Jiang:2015hri}. Around the critical point, anomalies, such as enhanced fluctuations are expected to appear \cite{Luo:2017faz}. Physical program of NA61/SHINE at SPS \cite{Gazdzicki:2008pu, Gazdzicki:2011fx} and MPD at NICA \cite{Toneev:2007yu} is focused on these investigations.
Correlations and fluctuations are also studied at LHC energy \cite{Abelev:2012pv,ATLAS:2012as, Abelev:2009ag}
because they are sensitive to the very first stages of the collision \cite{Tribedy:2017yxb}. The so-called long-range correlations \cite{Altsybeev:2016uci} between observables in separated rapidity intervals play a special role. Causality implies that they are developed in the initial state.
In addition, a large enough rapidity gap can suppress the background contribution of resonance decays.

In the present work, the string-parton Monte Carlo model \cite{Kovalenko:2012ye, Kovalenko:2012nt, Kovalenko:2014tca} is applied to the relativistic nuclear collisions at high energy to calculate long-range correlations and fluctuations. The model includes the string fusion \cite{Braun:1992ss, Braun:1991dg, Amelin:1993cs, Braun:1997ch} effects and the finite rapidity length of strings, and also charge conservation in string fragmentation. We have calculated several types of strongly intensive event-by-event variables and long-range correlations for the experiment.

\section{Model description}
\label{model}

The present model \cite{Kovalenko:2012ye, Kovalenko:2012nt} describes nucleon-nucleon collisions at the partonic level. For the nuclear density of $Pb$ and $Xe$ the Woods-Saxon distribution of nucleons is taken:
\begin{equation}
\rho(r)=\dfrac{\rho_0}{1+\exp[(r-R)/d]},\end{equation}
$ \text{\newline with }R=6.62 \text{ fm, }  d=0.546 \text{ fm for } Pb;  R=5.42 \text{ fm, }d=0.57 \text{ fm } \text{for } Xe.$

Each nucleon contains a valence quark-diquark pair and a certain number of sea quark-antiquark pairs, distributed around the center of nucleon according to two-dimensional Gauss distributions. The number of sea pairs is distributed according to Poisson law. The combined exclusive distribution of parton coordinates in the transverse plane ($r_j$) preserves the conservation law for the center of energy \citep{Boyer_2005}:
\begin{equation}
\sum\limits_{j=1}^N \vec r_j \cdot x_j =0,
\end{equation}
where $x_j$ is a longitudinal momentum fraction of parton $j$. These variables follow the exclusive distribution, which for an arbitrary number of quark-antiquark pairs has the following form:
\begin{equation}
			\rho(x_1,... x_N)=c\cdot\prod\limits_{j=1}^{N-1} x_j^{-\frac{1}{2}} 
							\cdot x_N^{\alpha_N} \delta(\sum\limits_{i=1}^N x_i -1).
			\end{equation}

The valence quark is labeled by N-1, the diquark -- by N, and the other indexes refer to sea quarks and antiquarks.			
${\alpha_N}$ = 3/2 (ud-diquark), ${\alpha_N}$ = 5/2 (uu-diquark).
 This distribution corresponds to the inclusive distributions, used in the model with quark-gluon strings \cite{Kaidalov:1985jg,Arakelian:2002iw}.
Corresponding algorithms for sampling with the exclusive distributions mentioned above are discussed in \cite{Kovalenko:2012nt}.

An elementary interaction is realized in the model of colour dipoles \cite{Flensburg:2008ag, Gustafson:2009qz}. Taking also into account also the confinement scale ($r_{max}$), the probability amplitude for a scattering the dipoles with transverse coordinates  $(\vec r_1, \vec r_2)$ and $(\vec r_3, \vec r_4)$  is given by:
		\begin{equation} \label{newformula}
			f=\frac{\alpha_s^2}{2}\left[ K_0\left(\frac{|\vec r_1-\vec r_3|}{r_{max}}\right) +
			K_0\left(\frac{|\vec r_2-\vec r_4|}{r_{max}}\right) 
			- K_0\left(\frac{|\vec r_1-\vec r_4|}{r_{max}}\right)
			- K_0\left(\frac{|\vec r_2-\vec r_3|}{r_{max}}\right)	\right]^2
		\end{equation}	
		Here  $\alpha_S$ is assumed to be an effective coupling constant and treated as a model parameter. 	The total probability of inelastic interaction of two protons in the eikonal form appears as ${p=1-e^{-\sum\limits_{i,j} f_{ij}}}$,
where the summation is done over all the dipoles.

Then it is assumed, that if there is a collision between two dipoles,
 two quark-gluon strings
are stretched between the ends of the dipoles,
and the process of string fragmentation gives
observable particles. The string in the rapidity space is formed with rapidity ends $y_{\text{min}}$ and $y_{\text{max}}$ \cite{Kovalenko:2012nt, Kovalenko:2014tca}, which emits the produced particles uniformly in rapidity between the string  ends with some mean number of charged particles per rapidity $\mu_0$. The emission happens independently in each rapidity interval, with Poisson distribution. 
Each parton can interact with another one only once, forming a pair of quark-gluon strings, hence, producing particles.
In the model, the energy-momentum of the nucleons is conserved on the partonic level if we take into account both interacting and non-interacting partons. It is also fully valid at the level of strings, however during the production of particles it is approximate.

The transverse position of a string is assigned to the arithmetic mean of the transverse coordinates of the partons at the ends of the string.
Due to the finite transverse size of the strings they overlap, that in the framework of the string fusion model
 \cite{Braun:1992ss, Braun:1991dg, Amelin:1993cs, Braun:1997ch}  gives a source with higher tension.
The mean multiplicity of charged particles and mean $p_t$ originated from the cell where $k$ strings are overlapping look as follows:
\[
	\left\langle \mu\right\rangle_k=\mu_0 \sqrt{k}, \hspace*{1cm}
	\left\langle p_t\right\rangle_k=p_0 \sqrt[4]{k}.
\]
Here $\mu_0$ and $p_0$ are the mean charged multiplicity from one single string per rapidity unit
and mean transverse momentum from one single string. 
For the overlapping of the strings of a finite rapidity length, the rapidity space is divided into the regions where the number of overlapped strings is an integer, and they are processed separately.
All parameters of the model are constrained from the data on inelastic cross section and multiplicity in pp, p-Pb and Pb-Pb collisions \cite{Kovalenko:2014tca}.

The charge differentiation of the particles in this model is performed in the following way.
If we neglect the production of the double charge at string fragmentation, each emitted particle can change the charge going along rapidity only by 0 or 1 (emitting of neutral or charged pion). After removing the neutral particles, we see that the charge of particles emitted from one string (or a cluster of fused strings) must alternate between 1 to -1 along the rapidity axis. This procedure is fulfilled for each string cluster. 

\section{Definition of observables}

Firstly, we define the correlation coefficient of long range correlations between observables in two separated rapidity windows as follows:

\begin{equation}
b=\dfrac{\langle B F \rangle - \langle B \rangle \langle F \rangle}{   \langle F^2 \rangle -  \langle F \rangle^2},
\end{equation}
where each of ($B$, $F$) could be either $N_{\mathrm{ch}}$ -- the number of charged particles in the rapidity window or $p_t$ -- the mean event transverse
momentum of charged particles in the given window:
\begin{equation}
p_t = \dfrac{1}{N_{\mathrm{ch}}} \sum\limits_{i=1}^{N_{\mathrm{ch}}}{{p_t}_i}.
\end{equation}

Correspondingly, three types of correlation coefficients can be studied: the n-n,  pt-n,  pt-pt correlations.

Unfortunately, the n-n and pt-n correlation coefficients are shown to be very dependent of the centrality class width and details of the centrality determination method \cite{Kovalenko:2013jya} (note pt-pt has not this property). 
To overcome this dependence, the so-called strongly intensive variables can be used	\cite{Gorenstein:2011vq, Gazdzicki:2013ana, Andronov:2015bqn, Andronov:2018xom}.
In this paper, we apply only $\Sigma[F,B]$ defined as:
\begin{equation}
\label{SigmaFB}
\Sigma(F,B)\equiv\frac{\av{F}\,\omega_B+\av{B}\,\omega_F-2\, \left(\av{FB}-\av{F} \av{B} \right) }{\av{F} +\av{B}}  \ ,
\end{equation}
where $\omega_F \equiv  \dfrac{\av{F^2}-\av{F^2}}{\av{F}}  \ , \omega_B =  \dfrac{\av{B^2}-\av{B^2}}{\av{B}}  \ $ -- scaled (or relative) variance.

Variables $F$ and $B$ here must be extensive. For them, we have used  the total multiplicity in a rapidity window ($N$), and total transverse momentum ($P_T$).

To study the charge fluctuations, we have used dynamical fluctuation measure, $\nu_{dyn}$, defined as follows:
\begin{equation}
\nu_{dyn}=\left\langle \left( \dfrac{N_+}{\langle N_+ \rangle} - \dfrac{N_-}{\langle N_- \rangle}  \right)^2  \right\rangle  - \left( \dfrac{1}{\langle{N_+}\rangle} + \dfrac{1}{\langle N_- \rangle} \right).
\end{equation}
It can be shown that the variable $N_{ch} \nu_{dyn} = \langle{N_+} + {N_-}\rangle\nu_{dyn}$ is also strongly intensive.

The increase and non-monotonic behavior of the mentioned strongly intensive variables can be an indication of the critical point of the QCD phase transition \cite{Stephanov:2004wx}, so they are used in the experimental investigations \cite{Aduszkiewicz:2017mei, Andronov:2018ukf, Andronov:2018ccl}.

\section{Results}
\label{results}

\begin{figure}[h]
\centering
\includegraphics[width=7cm,clip]{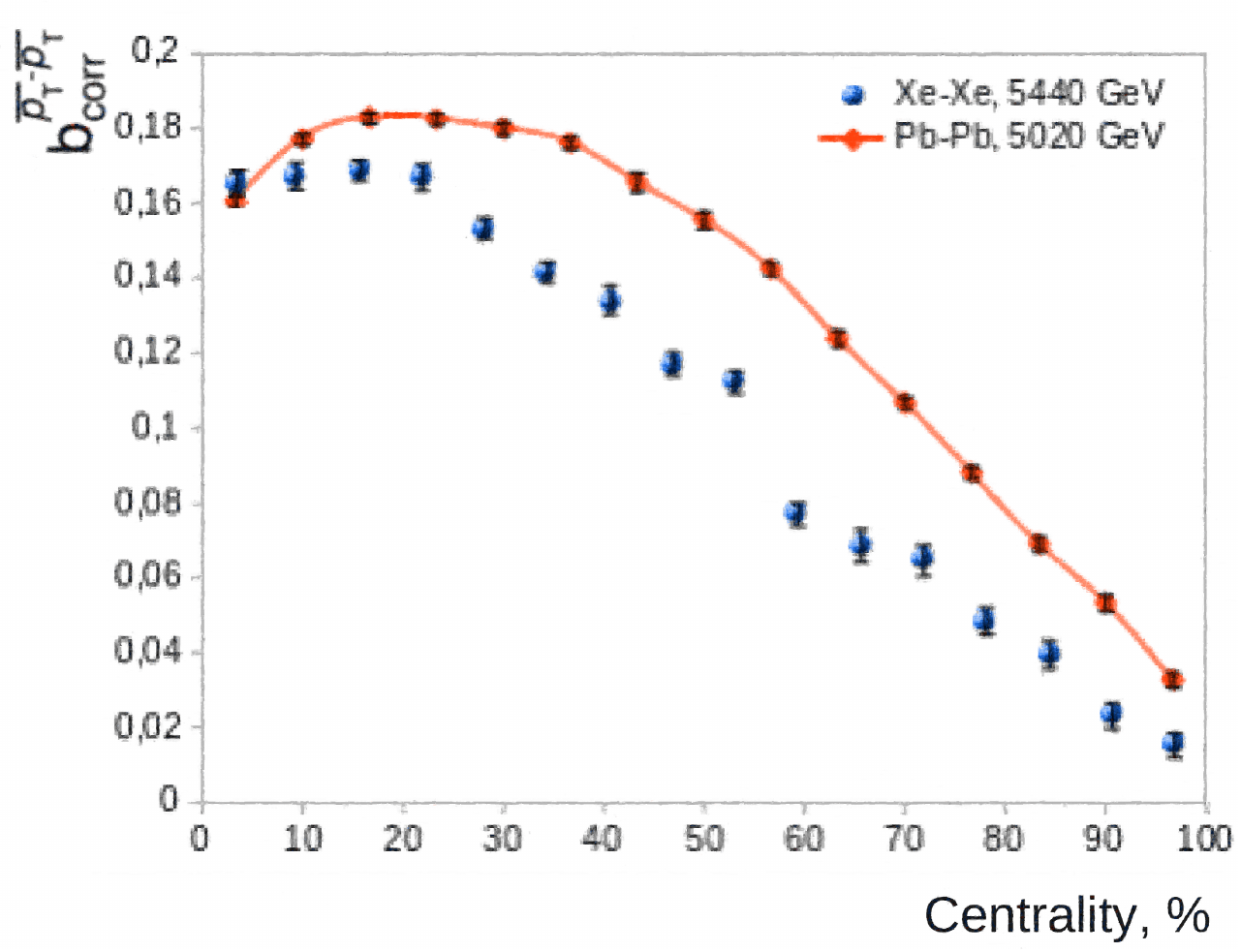}
\caption{Mean transverse momentum correlation coefficient as a function of centrality at LHC energy, calculated in the Monte Carlo model with string fusion for Pb-Pb and Xe-Xe nuclei.}
\label{fig-1}       
\end{figure}

Figure \ref{fig-1} shows the centrality dependence of the mean transverse momentum correlation coefficient.
We have calculated it for two colliding systems -- Pb-Pb and Xe-Xe at the LHC energy to demonstrate also the system size dependence of the pt-pt correlation coefficient.
In case of Pb-Pb collisions, the model predicts \cite{Kovalenko:2013jya} non-monotonic behaviour with centrality, which was also confirmed by experimental data
\cite{Altsybeev:2017eoj}, but it is hard to reproduce in the models \cite{Kovalenko:2016bcx}. The position of the peak can be attributed to the point of the saturation of the color field.

In the case of a smaller system, the saturation region seems not to be reached because even in most central Xe+Xe collisions the required density is not achieved. Similar behavior has been also obtained in p-Pb collisions \cite{Kovalenko:2014ika}.

\begin{figure}[h]
\centering
\includegraphics[width=7cm,clip]{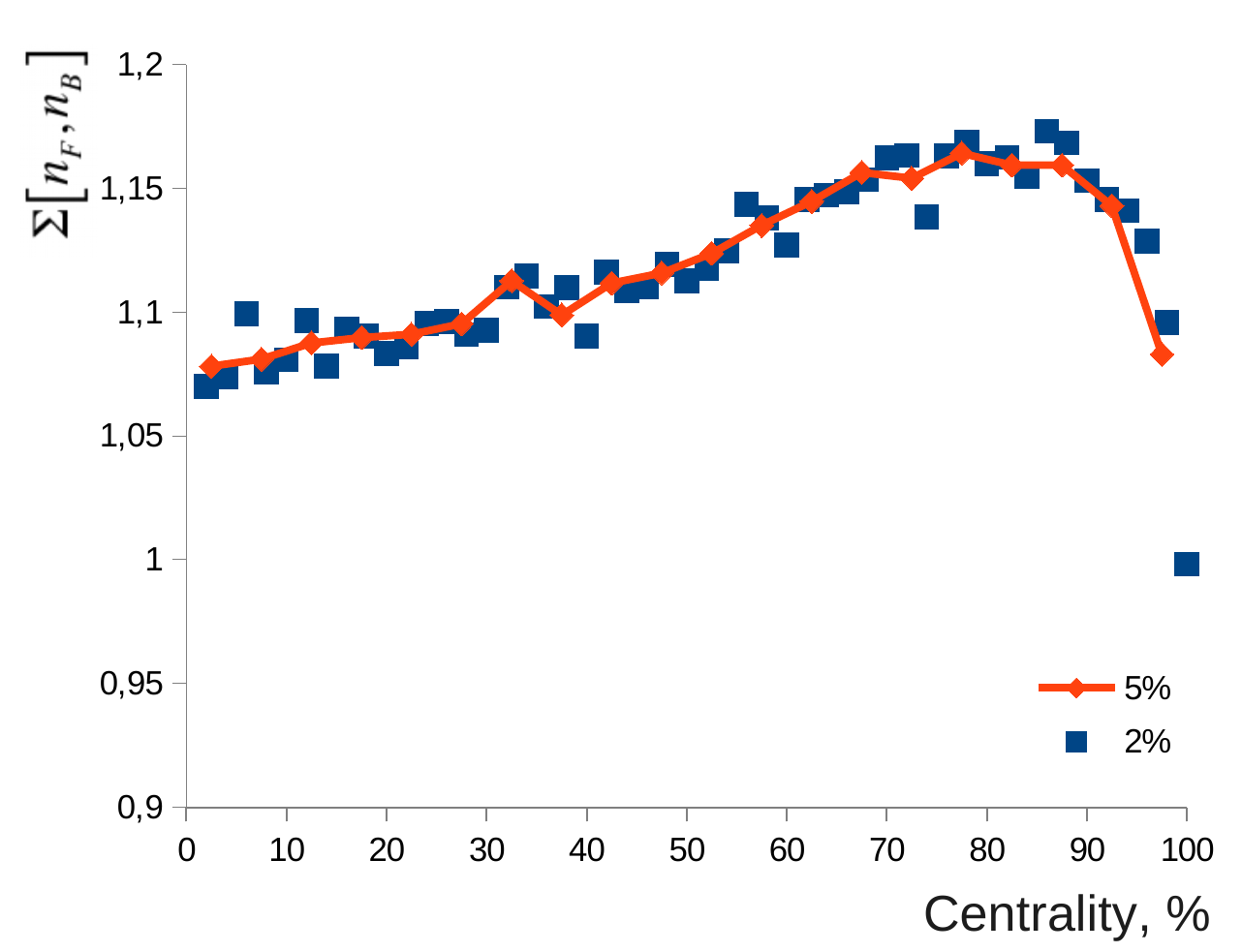}
\caption{Strongly intensive quantity $\Sigma_{NN}$ as a function of centrality, calculated in the Monte Carlo model with string fusion for Pb-Pb collisions at 2.76 TeV. Two centrality class binning (2\% and 5\%) are shown by different markers.}
\label{fig-2}       
\end{figure}

In figure \ref{fig-2} the forward-backward strongly intensive measure $\Sigma$ is shown for sufficiently large separated rapidity intervals (-3.8, -3), (3, 3.8). We see that above the very peripheral collisions the value of $\Sigma$ decreases with centrality.
This behavior has been recently observed in the experimental data; however other models cannot reproduce this effect \cite{IwonaHotQuarks,Sputowska:2016vmz}.

\begin{figure}[h]
\centering
\includegraphics[width=6.2cm,clip]{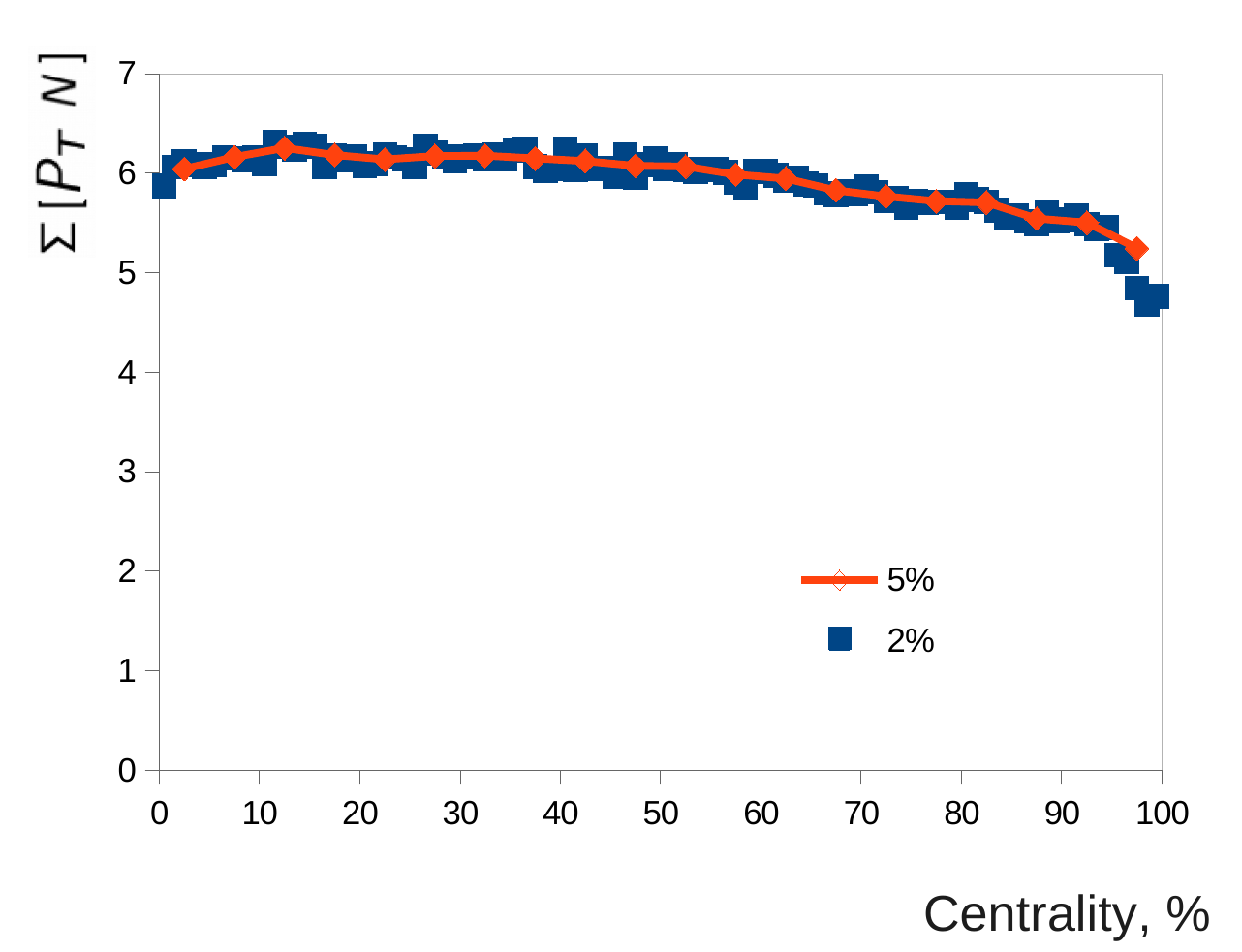}
\includegraphics[width=6.2cm,clip]{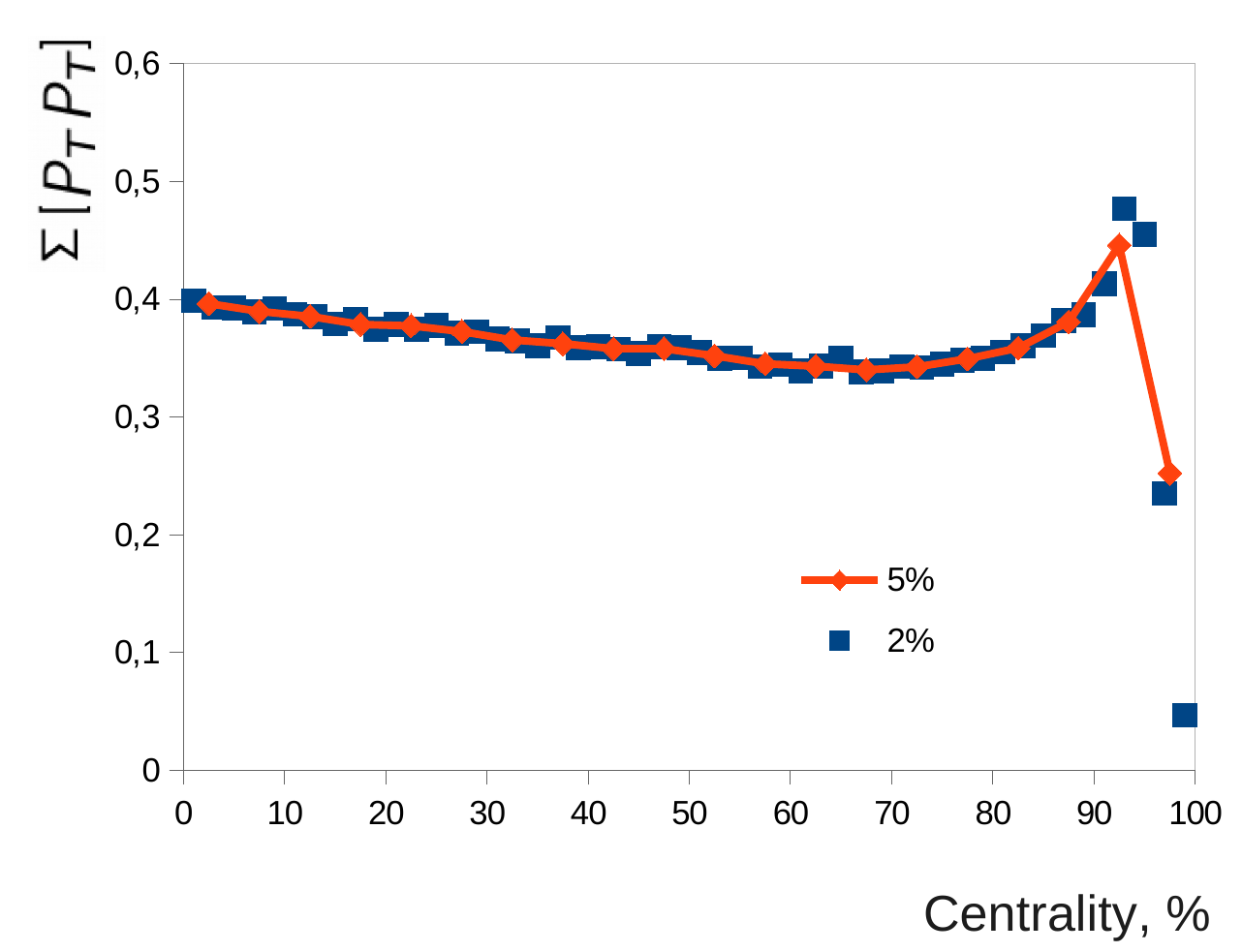}
\caption{Strongly intensive quantities $\Sigma_{P_{T}N}$ (left) and $\Sigma_{P_T P_T}$ (right) as a function of centrality, calculated in the Monte Carlo model with string fusion for Pb-Pb collisions at 2.76 TeV. As in fig. \ref{fig-2}, two centrality class binning (2\% and 5\%) are shown by different markers.}
\label{fig-3}       
\end{figure}

Figure \ref{fig-3} shows the centrality dependence of the two other types of the strongly intensive quantities between the observables in  the separated rapidity intervals, namely, $\Sigma[P_T, N]$ and $\Sigma[P_T,P_T]$.

 $\Sigma[P_T, N]$ is rather flat, what reminds one the dependence of the average transverse momentum with multiplicity \cite{Abelev:2013bla}. In pp and p-Pb collisions the latter is less trivial so it is worth checking the multiplicity dependence of a strongly intensive pt-n measure there.

The strongly intensive transverse momentum variable $\Sigma[P_T,P_T]$  (right plot of figure \ref{fig-3}) shows, in particular, the non-monotonic behavior. This reminds us the behavior of $b_{p_t-p_t}$. However, the position of the minimum here is at a different point than the maximum of $b_{p_t-p_t}$ (here it lies in the range of a less central collision). 

The maxima appearing in figures \ref{fig-2} and \ref{fig-3} in the region of very peripheral collisions could be related with the transition into a region of diffusive edges of nuclei. But, this centrality range is hardly accessible by the experiment \cite{Abelev:2013qoq}.

To summarize, all the three variants of the forward-backward strongly intensive measure indeed does not depend on the centrality class width; they provide additional non-trivial information which can not be extracted only from the correlation coefficient and thus should be widely experimentally studied, at LHC energy and also at lower such as MPD at NICA.

\begin{figure}[h]
\centering
\includegraphics[width=12cm,clip]{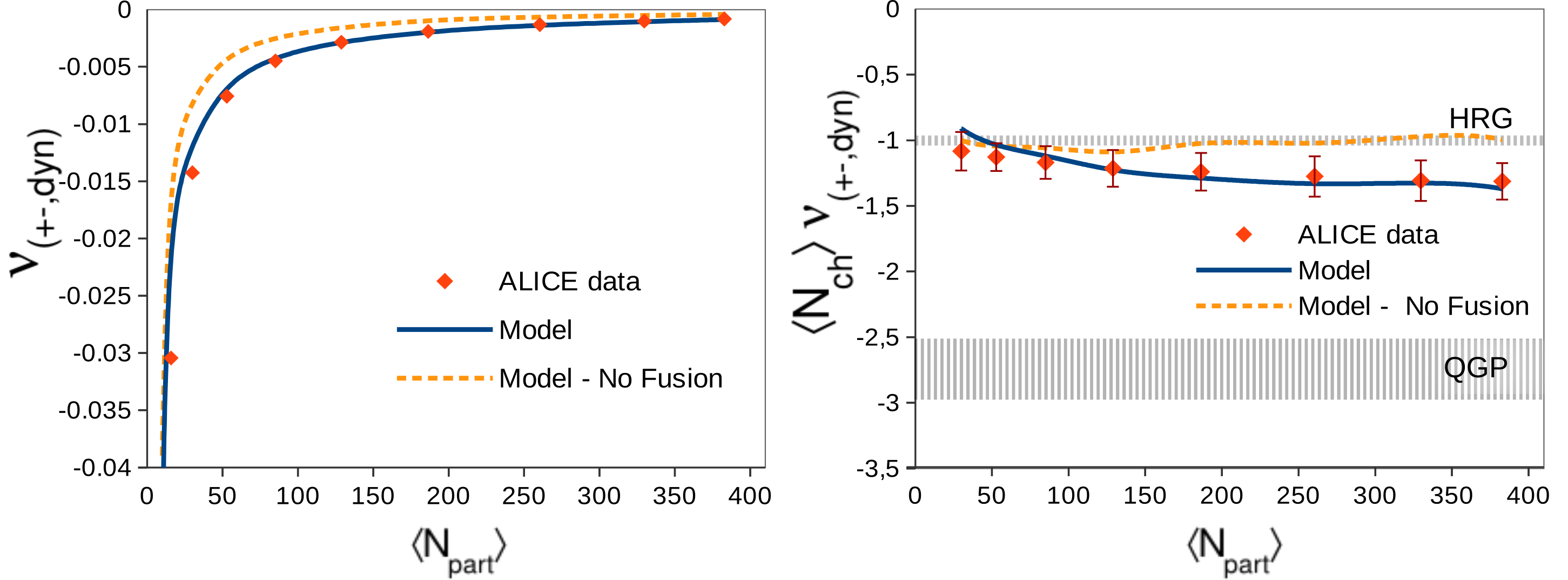}
\caption{Dynamical charge fluctuation $\nu_{dyn}$ as a function of centrality (number of participant nucleons) for Pb-Pb collisions at 2.76 TeV, calculated in the Monte Carlo model with string fusion and compared with experimental data \cite{Abelev:2012pv}. 
The solid line corresponds to the model with string fusion, dashed -- without fusion.
The right panel shows $\nu_{dyn}$ multiplied by multiplicity. Areas that correspond to hadron resonance gas (HRG) and quark-gluon plasma (QGP) are  hatched. }
\label{fig-4}       
\end{figure}

The figure \ref{fig-4} shows the charge fluctuations in terms of $\nu_{dyn}$ (left plot) and $N_{ch} \nu_{dyn}$ (right plot) with the comparison to the experimental data \cite{Abelev:2012pv}. The results have shown that taking into account the string fusion improves the agreement with the experimental data. Without fusion, the model behaves as hadron resonance gas, wheres an adding of a string collectivity modifies $N_{ch} \nu_{dyn}$ and makes it centrality-dependent.

%

%
%
%

\section{Conclusions}

Various strongly intense quantities are studied in AA collisions at the LHC energy, depending on the centrality compared (where possible) with experimental data  of ALICE.
All analyzed strongly intensive variables and correlations between intensive observables have shown robust behavior against the volume fluctuations and the details of centrality determination.

 It is shown that the string fusion model correctly qualitatively describes the $\Sigma_{NN}$ dependence  on the centrality. Charge fluctuations in Pb-Pb collisions at the energy of 2.76 TeV have been investigated. The correct description of the experimental ALICE data for $\nu_{dyn}$ has been obtained depending on the centrality of the collisions (string fusion effects improve the agreement with the experiment). The string fusion model explains the non-monotonic behavior of pt-pt forward-backward correlations in Pb-Pb collisions and predicts that at a smaller system this should not be observed.

 The studied observables provide a reference to the properties of the initial state of AA collisions and are useful to study both at the LHC and NICA energies.

\section{Acknowledgments}

The research was supported by the grant of the Russian Foundation for Basic Research (project 18-32-01055 mol\_a).

%
%

\bibliography{Baldin2018.bib}

\begin{thebibliography}{38}

\bibitem{Jiang:2015hri}
L.~Jiang, P.~Li, H.~Song, Phys. Rev. C \textbf{94}, 024918 (2016),
  arXiv:~\texttt{1512.06164}

\bibitem{Luo:2017faz}
X.~Luo, N.~Xu, Nucl. Sci. Tech. \textbf{28}, 112 (2017), arXiv:~\texttt{1701.02105}

\bibitem{Gazdzicki:2008pu}
M.~Gazdzicki, Eur. Phys. J. CST \textbf{155}, 37 (2008), arXiv:~\texttt{0801.4919}

\bibitem{Gazdzicki:2011fx}
M.~Gazdzicki (NA49, NA61/SHINE Collaborations), J.Phys.G \textbf{38}, 124024
  (2011), arXiv:~\texttt{1107.2345}

\bibitem{Toneev:2007yu}
V.~Toneev, PoS CPOD\textbf{07}, 057 (2007), arXiv:~\texttt{0709.1459}

\bibitem{Abelev:2012pv}
B.~Abelev et~al. (ALICE Collaboration), Phys. Rev. Lett. \textbf{110}, 152301
  (2013), arXiv:~\texttt{1207.6068}

\bibitem{ATLAS:2012as}
G.~Aad et~al. (ATLAS Collaboration), JHEP \textbf{1207}, 019 (2012),
  arXiv:~\texttt{1203.3100}

\bibitem{Abelev:2009ag}
B.~Abelev et~al. (STAR Collaboration), Phys. Rev. Lett. \textbf{103}, 172301
  (2009), arXiv:~\texttt{0905.0237}

\bibitem{Tribedy:2017yxb}
P.~Tribedy, J. Phys. Conf. Ser. \textbf{832}, 012013 (2017)

\bibitem{Altsybeev:2016uci}
I.~Altsybeev, J. Phys. Conf. Ser. \textbf{798}, 012056 (2017),
  arXiv:~\texttt{1611.10090}

\bibitem{Kovalenko:2012ye}
V.~Kovalenko, V.~Vechernin, PoS Baldin ISHEPP \textbf{2012}, 077 (2012),
  arXiv:~\texttt{1212.2590}

\bibitem{Kovalenko:2012nt}
V.~Kovalenko, Phys. Atom. Nucl. \textbf{76}, 1189 (2013), arXiv:~\texttt{1211.6209}

\bibitem{Kovalenko:2014tca}
V.~Kovalenko, PoS QFTHEP\textbf{2013}, 052 (2013)

\bibitem{Braun:1992ss}
M.~Braun, C.~Pajares, Phys. Lett. B \textbf{287}, 154 (1992)

\bibitem{Braun:1991dg}
M.~Braun, C.~Pajares, Nucl. Phys. B \textbf{390}, 542 (1993)

\bibitem{Amelin:1993cs}
N.~Amelin, M.~Braun, C.~Pajares, Phys. Lett. B \textbf{306}, 312 (1993)

\bibitem{Braun:1997ch}
M.~Braun, C.~Pajares, J.~Ranft, Int. J. Mod. Phys. A \textbf{14}, 2689 (1999),
  arXiv:~\texttt{hep-ph/9707363}

\bibitem{Boyer_2005}
T.H. Boyer, American Journal of Physics \textbf{73}, 953 (2005)

\bibitem{Kaidalov:1985jg}
A.~Kaidalov, O.~Piskunova, Z. Phys. C \textbf{30}, 145 (1986)

\bibitem{Arakelian:2002iw}
G.~Arakelian, A.~Capella, A.~Kaidalov, Y.~Shabelski, Eur. Phys. J. C
  \textbf{26}, 81 (2002)

\bibitem{Flensburg:2008ag}
C.~Flensburg, G.~Gustafson, L.~Lonnblad, Eur. Phys. J. C \textbf{60}, 233
  (2009), arXiv:~\texttt{0807.0325}

\bibitem{Gustafson:2009qz}
G.~Gustafson, Acta Phys. Polon. B \textbf{40}, 1981 (2009), arXiv:~\texttt{0905.2492}

\bibitem{Kovalenko:2013jya}
V.~Kovalenko, V.~Vechernin, EPJ Web Conf. \textbf{66}, 04015 (2014),
  arXiv:~\texttt{1308.6618}

\bibitem{Gorenstein:2011vq}
M.I. Gorenstein, M.~Gazdzicki, Phys. Rev. C \textbf{84}, 014904 (2011),
  arXiv:~\texttt{1101.4865}

\bibitem{Gazdzicki:2013ana}
M.~Gazdzicki, M.~Gorenstein, M.~Mackowiak-Pawlowska, Phys. Rev. C \textbf{88},
  024907 (2013), arXiv:~\texttt{1303.0871}

\bibitem{Andronov:2015bqn}
E.V. Andronov, Theor. Math. Phys. \textbf{185}, 1383 (2015), [Teor. Mat. Fiz.
  185, no. 1, 28 (2015)]

\bibitem{Andronov:2018xom}
E.~Andronov, V.~Vechernin (2018), Eur. Phys. J. \textbf{A} 55, 14 (2019).  arXiv:~\texttt{1808.09770}

\bibitem{Stephanov:2004wx}
M.A. Stephanov, Prog. Theor. Phys. Suppl. \textbf{153}, 139 (2004), [Int. J.
  Mod. Phys.A20,4387(2005)], arXiv:~\texttt{hep-ph/0402115}

\bibitem{Aduszkiewicz:2017mei}
A.~Aduszkiewicz, Nucl. Phys. A \textbf{967}, 35 (2017), arXiv:~\texttt{1704.08071}

\bibitem{Andronov:2018ukf}
E.~Andronov, EPJ Web of Conf. \textbf{191}, 05002 (2018)

\bibitem{Andronov:2018ccl}
E.~Andronov,  Nucl. Phys. A \textbf{982}, 835 (2019), arXiv: 1807.10737 [nucl-ex]  (2018)

\bibitem{Altsybeev:2017eoj}
I.~Altsybeev, KnE Energ. Phys. \textbf{3}, 304 (2018), arXiv:~\texttt{1711.04844}

\bibitem{Kovalenko:2016bcx}
V.~Kovalenko, V.~Vechernin, J. Phys. Conf. Ser. \textbf{798}, 012053 (2017),
  arXiv:~\texttt{1611.07274}

\bibitem{Kovalenko:2014ika}
V.~Kovalenko, AIP Conf. Proc. \textbf{1606}, 174 (2014), arXiv:~\texttt{1402.0281}

\bibitem{IwonaHotQuarks}
{I. A. Sputowska (For ALICE Collaboration)}, \emph{{Forward-backward
  correlations and multiplicity fluctuations in Pb-Pb collisions at
  $\sqrt{s_{NN}}$ = 2.76 TeV from ALICE at the LHC}} (Hot Quarks 2018.
  \url{https://indico.cern.ch/event/703015/contributions/3095137/}, 2018)

\bibitem{Sputowska:2016vmz}
I.~Sputowska, \emph{{Correlations in Particle Production in Nuclear Collisions
  at LHC Energies. \rm{Ph.D. thises.}}} (Cracow, INP, 2016)

\bibitem{Abelev:2013bla}
B.~Abelev et~al. (ALICE Collaboration), Phys. Lett. B \textbf{727}, 371 (2013),
  arXiv:~\texttt{1307.1094}

\bibitem{Abelev:2013qoq}
B.~Abelev et~al. (ALICE Collaboration), Phys. Rev. C \textbf{88}, 044909
  (2013), arXiv:~\texttt{1301.4361}

\end{thebibliography}

\end{document}